\documentclass[letter]{jpsj2} 
\title{Upper critical fields of the 11-system iron-chalcogenide superconductor FeSe$_{0.25}$Te$_{0.75}$}

\author{
Takanori \textsc{Kida}$^{1,4}$\thanks{E-mail: kida@mag.cqst.osaka-u.ac.jp}, 
Takahiro \textsc{Matsunaga}$^{1}$, 
Masayuki \textsc{Hagiwara}$^{1,4}$, \\
Yoshikazu \textsc{Mizuguchi}$^{2,3,4}$, 
Yoshihiko \textsc{Takano}$^{2,3,4}$, 
and Koichi \textsc{Kindo}$^{5}$
}

\inst{$^{1}$KYOKUGEN, Osaka University, 1-3 Machikaneyama, Toyonaka, Osaka 560-8531, Japan\\
$^{2}$National Institute for Materials Science, 1-2-1 Sengen, Tsukuba, Ibaraki 305-0047, Japan\\
$^{3}$University of Tsukuba, 1-1-1 Tennodai, Tsukuba, Ibaraki 305-8577, Japan\\
$^{4}$JST, Transformative Research-Project on Iron Pnictides (TRIP), Chiyoda-ku, Tokyo 102-0075, Japan\\
$^{5}$ISSP, The University of Tokyo, 5-1-5 Kashiwanoha, Kashiwa, Chiba 277-8581, Japan
}

\abst{
We have performed electrical resistivity measurements of a polycrystalline sample of FeSe$_{0.25}$Te$_{0.75}$, 
which exhibits superconductivity at $T_{\rm c} \sim 14$~K, 
in magnetic fields up to 55~T to determine the upper critical field $\mu_{0}H_{\rm c2}$. 
In this compound, very large slopes of $\mu_{0}H_{\rm c2}$ 
at the onset, the mid-point, the zero-resistivity temperatures on superconductivity are determined to be 
$-13.7$, $-10.1$, and $-6.9$~T/K, respectively. 
The observed $\mu_{0}H_{\rm c2}(T)$s of this compound are considerably smaller than those expected from the Werthamer-Helfand-Hohenberg model, manifesting the Pauli limiting behavior. 
These results suggest that this compound has a large Maki parameter, 
but it is smaller than that calculated for a weak-coupling superconductor, 
indicating a large superconducting gap of this compound as a strong-coupling superconductor. 
}

\kword{iron-based superconductor, FeSe, magnetoresistance, upper critical field }

\begin{document}
\maketitle

Recent discovery of superconductivity at $T_{\rm c}=26$~K in the iron-based LaFeAsO$_{1-x}$F$_{x}$~\cite{Kamihara} 
(abbreviated as the 1111-system) has attracted a considerable attention of condensed matter scientists due to an unusual interplay of superconductivity and magnetism arising from ubiquitous magnetic element Fe.  
In general, 
it has been believed that 
materials containing the magnetic elements are difficult to occur the superconductivity. 
By substitution of La atoms for other lanthanoid atoms, 
$T_{\rm c}$ increases with increasing lanthanoid atomic number and shows a maximum value of 
$T_{\rm c}=55$~K for Sm atoms. 
As well as the CuO$_{2}$ layers in the high-$T_{\rm c}$ cuprate superconductors, 
the Fe-$Pn$ ($Pn$~=~P, As) layers in iron oxypnictides are responsible for the superconductivity, 
and the $Ln$-O ($Ln$~=~lanthanoid) layers provide charge carriers. 
The electrical structure of a high-$T_{\rm c}$ cuprate superconductor has been explained by 
the single band model, but it is difficult to understand the mechanism of its superconductivity 
due to the strongly electron correlation effects. 
While the electron correlation on an iron-based superconductor is thought to be much weak, 
we have to consider the multi-band effects of Fe $3d$ orbitals. 
Thus, the mechanism of superconductivity on an iron pnictide must be complicated. 

So far, several other groups of iron-based superconductors have been discovered, 
such as $Ae$Fe$_{2}$As$_{2}$ (abbreviated as the 122-system, $Ae$~=~alkali earth metals)~\cite{Rotter}, 
LiFeAs (abbreviated as the 111-system)~\cite{Pitcher}, 
tetragonal Fe$Ch$ (abbreviated as the 11-system, $Ch$~=~chalcogenides)~\cite{Hsu}, and 
($M_{2}Pn_{2}$)(Sr$_{4}$Sc$_{2}$O$_{6}$) (abbreviated as the Sc-22426 system, $M$~=~Fe, Ni)~\cite{Ogino}. 
In particular, the 11-system superconductors, such as FeSe$_{1-x}$~\cite{Hsu,Mizuguchi1}, FeSe$_{1-x}$Te$_{x}$~\cite{Yeh}, and FeTe$_{1-x}$S$_{x}$~\cite{Mizuguchi2}, are greatly important materials 
in understanding the mechanism of the superconductivity on the iron-based superconductors owing to their simple structure.

FeSe$_{1-x}$, which exhibits superconductivity at $T_{\rm c}=8$~K, 
has a tetragonal PbO-type structure ($P_{4}/nmm$) composed of the stacked FeSe layers along 
the $c$-axis~\cite{Hsu}. 
The superconductivity on FeSe$_{1-x}$ is significantly affected by applied 
pressure~\cite{Mizuguchi1,Masaki} and chalcogenide substitutions~\cite{Yeh,Mizuguchi2}. 
Especially, applied pressure only up to 4.15~GPa drastically enhances its $T_{\rm c}$ to $\sim 37$~K (d~ln$T_{\rm c}$/d$P \sim 0.91$)~\cite{Masaki}. 
The pressure effect of the superconductivity on FeSe$_{1-x}$ is larger than that of the other iron-based 
superconductors, $e.g.$, 
$T_{\rm c}$ of LaFeAsO$_{1-x}$F$_{x}$ increases to 43~K under a high pressure of 4~GP 
(d~ln$T_{\rm c}$/d$P \sim 0.16$)~\cite{Takahashi}. 

For a profound understanding of the mechanism of superconductivity on the iron-based superconductors, 
it is important to study the upper critical field ($\mu_{0}H_{\rm c2}$) because the $\mu_{0}H_{\rm c2}$ provides information about such as the coherent length, the anisotropy, and the pair-breaking mechanism. 
Transport measurements of high-$T_{\rm c}$ cuprate superconductors in very high magnetic fields have brought about useful information on not only the $\mu_{0}H_{\rm c2}$ but also the nature in the vicinity of the quantum phase transition point~\cite{Ando}. 
In the present study, we focus on the Te-substituted iron-chalcogenide FeSe$_{0.25}$Te$_{0.75}$ 
and report the $\mu_{0}H_{\rm c2}$ as a function of temperature in this compound, 
which shows the highest value of the upper critical field among the 11-systems~\cite{Yeh}. 
The $\mu_{0}H_{\rm c2}$s of the 1111- and the 122-systems have been already studied in high magnetic fields~\cite{Hunte,Fuchs}. 
To our knowledge, however, 
high magnetic field properties of the 11-system have not been reported yet.

Polycrystalline samples of FeSe$_{0.25}$Te$_{0.75}$ were prepared using a solid state reaction method 
as described in Ref.~\ref{sample}. 
We prepared the sample  with a typical dimension of 
4.0$\times$0.4$\times$0.4~mm$^{3}$ for the electrical resistivity ($\rho$) measurements. 
The current ($I$) direction was parallel to the magnetic field ($H$) and the longitudinal direction of the sample. 
The temperature dependence of $\rho$ was measured using a conventional dc four-probe technique 
in dc magnetic fields up to 7~T with a SQUID magnetometer (MPMS-$XL7$, Quantum Design). 
The $\rho$ in pulsed magnetic fields up to 55~T was measured by utilizing a non-destructive pulsed 
magnet at the High Magnetic Field Laboratory, KYOKUGEN in Osaka University. 
The duration of the pulsed magnetic field was about 40~msec.

\begin{figure}[tp]
\begin{center}
\includegraphics[width=0.65\textwidth,keepaspectratio=true]{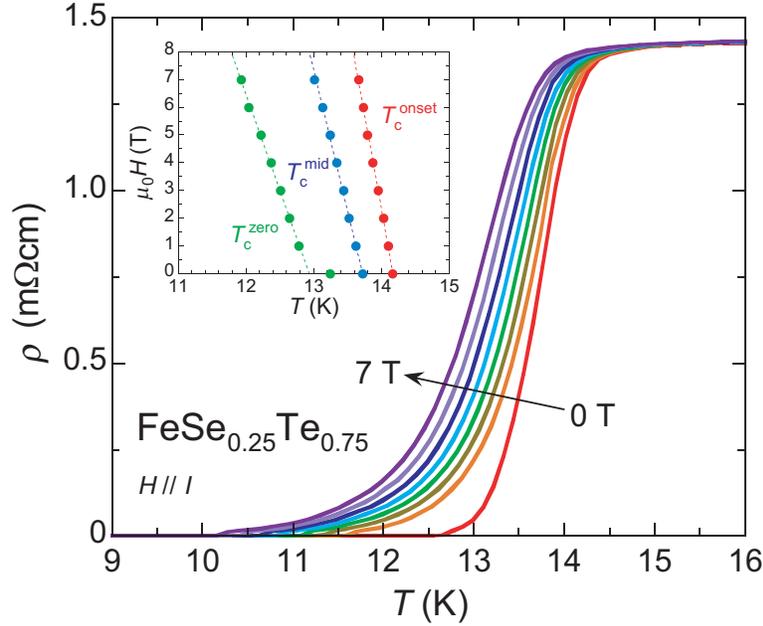}
\end{center}
\caption{(Color online)~Temperature dependence of the electrical resistivity of FeSe$_{0.25}$Te$_{0.75}$ 
in dc magnetic fields up to 7~T. 
The inset displays the temperature dependences of resistive upper critical field $\mu_{0}H_{\rm c2}(T)$ at three defined temperatures. 
The dashed lines are linear fits to the data.}
\label{fig01}
\end{figure}

Figure~\ref{fig01} shows the temperature dependence of $\rho$ of FeSe$_{0.25}$Te$_{0.75}$ 
in dc magnetic fields up to 7~T with an increment of 1~T ($H \parallel I$). 
We define three characteristic temperatures of the superconducting transition: 
the onset temperature $T_{\rm c}^{\rm onset}$ (90~\% of the normal state resistivity $\rho_{\rm n}(H, T)$), 
the mid-point temperature $T_{\rm c}^{\rm mid}$ (50~\% of $\rho_{\rm n}(H, T)$)), and 
the zero-resistivity temperature $T_{\rm c}^{\rm zero}$ (10~\% of $\rho_{\rm n}(H, T)$)) according to the definition reported in Ref.~\ref{Hunte}. 
The values of $T_{\rm c}$ at 0 T were determined to be $T_{\rm c}^{\rm onset}(0)=14.2$~K, 
$T_{\rm c}^{\rm mid}(0)=13.7$~K, and $T_{\rm c}^{\rm zero}(0)=13.2$~K, 
which are comparable to those reported before~\cite{Yeh}. 
The $T_{\rm c}$ decreases with increasing magnetic field. 
We plot the upper critical field ($\mu_{0}H_{\rm c2}$) at the above three defined temperatures in the inset of Fig.~\ref{fig01}. 
The curves of $\mu_{0}H$-$T_{\rm c}^{\rm onset}$ and $\mu_{0}H$-$T_{\rm c}^{\rm mid}$ are almost linear in temperature, but that of $\mu_{0}H$-$T_{\rm c}^{\rm zero}$ shows a slightly upturn curvature near 0 T. 
The slopes of $\mu_{0}H_{\rm c2}$ at $T_{\rm c}^{\rm onset}(0)$, $T_{\rm c}^{\rm mid}(0)$, and 
$T_{\rm c}^{\rm zero}(0)$ indicated by dashed lines in the inset of Fig.~\ref{fig01} are $-13.7$, $-10.1$, and $-6.9$~T/K, respectively. 
The slope at $T_{\rm c}^{\rm onset}(0)$ of this compound is steep against its $T_{\rm c}$ in comparison with those of the other FeAs-based superconductors~\cite{Fuchs}, $e.g.$, 
$-4.2$~T/K for LaFeAsO$_{0.93}$F$_{0.07}$ ($T_{\rm c}^{\rm onset}=25$~K) and 
$-6.3$~T/K for (Ba$_{0.55}$K$_{0.45}$)Fe$_{2}$As$_{2}$ ($T_{\rm c}^{\rm onset}=32$~K), as shown in Table~\ref{t1}. 
Based on the conventional one-band Werthamer-Helfand-Hohenberg (WHH) theory, which describes the orbital pair-breaking 
field (the zero-temperature upper critical field) of dirty type-II superconductors~\cite{WHH}, 
the $\mu_{0}H_{\rm c2}^{\rm orb}$ at 0 T is expressed by 
\begin{equation}
\mu_{0}H_{\rm c2}^{\rm orb}(0)=-0.69T_{\rm c}({\rm d}\mu_{0}H_{\rm c2}/{\rm d}T)|_{T_{\rm c}}. \label{eq1}
\end{equation}
Then, we can obtain $\mu_{0}H_{\rm c2}^{\rm onset}(0)=134$~T, 
$\mu_{0}H_{\rm c2}^{\rm mid}(0)=95.9$~T, and $\mu_{0}H_{\rm c2}^{\rm zero}(0)=62.1$~T. 

\begin{figure}[tp]
\begin{center}
\includegraphics[width=0.66\textwidth,keepaspectratio=true]{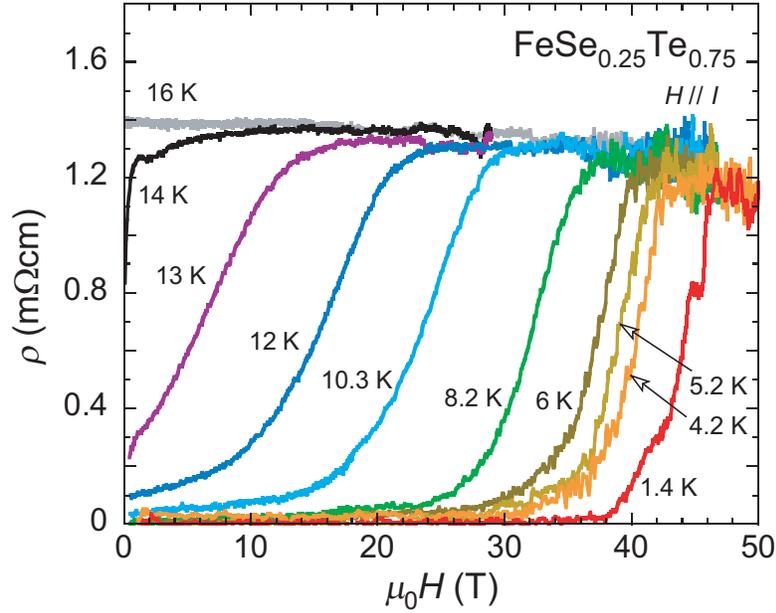}
\end{center}
\caption{(Color online)~Electrical resistivity $\rho(H)$ as a function of magnetic field up to 55~T at designated temperatures. The current direction is parallel to the field direction.}
\label{fig02}
\end{figure}

\begin{figure}[tp]
\begin{center}
\includegraphics[width=0.65\textwidth,keepaspectratio=true]{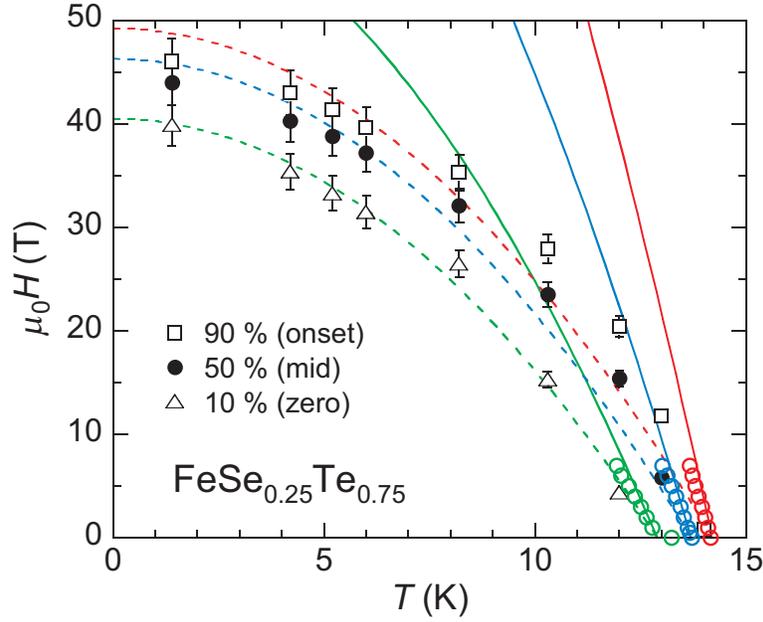}
\end{center}
\caption{(Color online)~Field-temperature ($H$-$T$) phase diagram for the polycrystalline sample of FeSe$_{0.25}$Te$_{0.75}$. 
Solid and dashed lines show the WHH model and the fitted ones by Eq.~(\ref{eq4}), respectively.}
\label{fig03}
\end{figure}

\begin{table}[tp]
\begin{center}
\caption{Upper critical field data of selected iron pnictides (onset) in Ref.~\ref{Fuchs} and the present sample (onset, mid-point, and zero-resistivity). $\mu_{0}H_{\rm c2}^{\rm orb}(0)$, $\mu_{0}H_{\rm p}(0)$, and $\mu_{0}H_{\rm c2}^{\rm p}(0)$ are the orbital (the WHH model), the Pauli, and the paramagnetically limited field, respectively. 
$\alpha_{0}$, $\alpha_{1}$ the Maki parameter ($\lambda_{\rm so}=0$) defined as $\alpha = \sqrt{2}H_{\rm c2}^{\rm orb}(0)/H_{\rm p}(0)$ and the fitted one, respectively. }\label{t1}
\vspace{0.5em}
\begin{tabular}{l|ccccccc} \hline
Compound & \multicolumn{1}{c}{$T_{\rm c}$} & \multicolumn{1}{c}{d$\mu_{0}H_{\rm c2}/$d$T$} & \multicolumn{1}{c}{$\mu_{0}H_{\rm c2}^{\rm orb}(0)$} & \multicolumn{1}{c}
{$\mu_{0}H_{\rm p}(0)$} & \multicolumn{1}{c}{$\mu_{0}H_{\rm c2}^{\rm p}(0)$} & \multicolumn{1}{c}{$\alpha_{0}$ } & \multicolumn{1}{c}{$\alpha_{1}$ } \\
 & \multicolumn{1}{c}{(K)} & \multicolumn{1}{c}{(T/K)} & \multicolumn{1}{c}{(T)} & \multicolumn{1}{c}{(T)} & \multicolumn{1}{c}{(T)} & &  \\ \hline
LaFeAsO$_{0.93}$F$_{0.07}$ & 25.0 & \llap{$-$}4.2 & 72 & 305 & 69 & 0.3 & -- \\
(Ba$_{0.55}$K$_{0.45}$)Fe$_{2}$As$_{2}$ & 32.0 & \llap{$-$}6.3 & 138 & 130 & 76  & 1.5 & -- \\ \hline
onset & 14.2 & \llap{$-$}13.7 & 134 & 26.4 & 18.4  & 7.2 & 2.5 \\
mid-point & 13.7 & \llap{$-$}10.1 & \phantom{0}95.9 & 25.5 & 17.8 & 5.3 & 1.8 \\
zero-resistivity & 13.2 & \phantom{0}\llap{$-$}6.9 & \phantom{0}62.1 & 24.0 & 16.2  & 3.7 & 1.4 \\ \hline
\end{tabular}
\end{center}
\end{table} %

The results of the high-field resistivity as a function of magnetic field ($H \parallel I$) 
in the temperature range from 1.4~K to 16~K are shown in Fig.~\ref{fig02}. 
Below $T_{\rm c}$, the superconductivity is suppressed by magnetic field. 
It is clear that the $\rho(H, T)$ curves are shifted to lower magnetic fields upon increasing temperature. 
The values of $\rho(H, T)$ in the normal state just above the critical field are 
nearly constant independent of temperature. 
Magnetoresistance $(\rho(H)-\rho(0))/\rho(0)$ in the normal state (above $T_{\rm c}$) is almost zero, 
as seen in the results of Fig.~\ref{fig01} above about 15 K. 
Similar results were also reported in other iron-based superconductors~\cite{Fuchs}. 
Here, we defined three characteristic fields: 
the high-field onset of superconducting transition $\mu_{0}H_{\rm c2}^{\rm onset}$ (90~\% of $\rho_{\rm n}$), 
the mid-point field $\mu_{0}H_{\rm c2}^{\rm mid}$ (50~\% of $\rho_{\rm n}$), and 
the zero-resistivite field $\mu_{0}H_{\rm c2}^{\rm zero}$ (10~\% of $\rho_{\rm n}$). 
At 1.4~K, the resistive critical fields are $\mu_{0}H_{\rm c2}^{\rm onset} \sim 46$~T, 
$\mu_{0}H_{\rm c2}^{\rm mid} \sim 44$~T, and $\mu_{0}H_{\rm c2}^{\rm zero} \sim 40$~T. 
Using the results of magnetic field and temperature dependences of $\rho$ in dc and pulsed magnetic fields, 
we illustrated the field-temperature ($\mu_{\rm 0}H$-$T$) phase diagram for the present sample as shown in Fig.~\ref{fig03}. 
The upper critical fields exhibit a saturation behavior at low temperature, 
as observed in most strongly disordered iron-based superconductors~\cite{Fuchs}. 
On the contrary, 
it was reported by Hunte $et~al$.~\cite{Hunte} that 
the upper critical fields of the 1111-system LaFeAsO$_{0.89}$F$_{0.11}$  exhibit upturn behavior, 
which is expected theoretically in dirty two-gap superconductors~\cite{Gurevich}. 
The extrapolated zero-temperature upper critical fields of the present sample are significantly 
smaller than those of the WHH model as indicated by the solid lines in Fig.~\ref{fig03}. 
The temperature dependence of the upper critical field empirically obeys the following relationship, 
$\mu_{0}H_{\rm c2}(T)/ \mu_{0}H_{\rm c2}(0) = 1-(T/T_{\rm c})^{2}$. 
Usually, the upper critical fields of the disordered superconductors deviate from the WHH curve 
and become smaller than those expected from the WHH model at low temperatures~\cite{Fuchs}. 

In the following, we discuss the suppression of the upper critical fields. 
In the standard BCS model, orbital effects limit the emergence of superconductivity and 
the superconductivity is destroyed when the kinetic energy of the charges exceeds the condensation energy of the Cooper pairs. 
The zero-temperature critical field $\mu_{0}H_{\rm c2}^{\rm orb}(0)$ expected from the WHH model is given by Eq.~(\ref{eq1}). 
This critical field must be suppressed by some reason. 
Most plausible cause is the Pauli spin susceptibility. 
The Pauli spin susceptibility energy plays an important role in suppressing the superconducting state, 
especially in some heavy fermion systems. 
In the Pauli limit or the paramagnetic effects (called also the spin Zeeman effects), 
superconductivity is destroyed when the polarization energy of the spins exceeds the condensation energy due to 
partial alignment of the spins. 
The zero-temperature Pauli limiting field for weakly coupled superconductors is given by 
\begin{equation}
\mu_{0}H_{\rm p}(0) = 1.86 T_{\rm c}\sqrt{1 + \lambda_{\rm so}} = 1.06 \Delta_{0}\sqrt{1 + \lambda_{\rm so}}, \label{eq2}
\end{equation}
where $\lambda_{\rm so}$ is the spin-orbit scattering constant, and $\Delta_{0}$ is the superconductivity gap 
(2$\Delta_{0}=3.52k_{\rm B}T_{\rm c}$ for the conventional weak-coupling superconductor). 
According to Maki~\cite{Maki}, the paramagnetically limited field is given by 
\begin{equation}
\mu_{0}H_{\rm c2}^{\rm p}(0)=\mu_{0}H_{\rm c2}^{\rm orb}(0)/\sqrt{1+\alpha^{2}}, \label{eq3}
\end{equation}
where $\alpha$ is the Maki parameter given by $\alpha = \sqrt{2}H_{\rm c2}^{\rm orb}(0)/H_{\rm p}(0)$, 
and $H_{\rm c2}^{\rm orb}(0)$ and $H_{\rm p}(0)$ are calculated from Eqs.~(\ref{eq1}) and (\ref{eq2}). 
It is obvious that a finite $\alpha$ suppresses the zero-temperature critical field $\mu_{0}H_{\rm c2}^{\rm p}(0)$. 
Accordingly, $H_{\rm c2}^{\rm p} < H_{\rm p} < H_{\rm c2}^{\rm orb}$ is fulfilled for $\alpha > \sqrt{2}$. 
For the present sample, 
the values of $\mu_{0}H_{\rm c2}^{\rm orb}(0)$, $\mu_{0}H_{\rm p}(0)$, $\mu_{0}H_{\rm c2}^{\rm p}(0)$, and 
$\alpha_{0}$ for $\lambda_{\rm so}=0$ are shown in Table~\ref{t1}. 
The onset value of $\mu_{0}H_{\rm p}(0)$ for the present sample is about $1/5$ of $\mu_{0}H_{\rm c2}^{\rm orb}(0)$, resulting in $\alpha_0$=7.2 which is much larger than that for other two Fe pnictide samples in Table~\ref{t1}. 

Within the WHH approach, the $\mu_{0}H_{\rm c2}(T)$ curve depends sensitively on the magnitude of the Maki 
parameter. 
The suppression of the $\mu_{0}H_{\rm c2}(T)$ near 0~K is predicted with increasing $\alpha$. 
Since the $\alpha_{0}$ is calculated by using $T_{\rm c}$ and ${\rm d}\mu_{0}H_{\rm c2}/{\rm d}T$ at $T_{\rm c}$, 
it is difficult to estimate the $\mu_{0}H_{\rm c2}^{\rm p}(0)$ from Eq.~\ref{eq3} accurately. 
Thus, 
under the assumption of $\lambda_{\rm so}=0$, we estimate experimental values of the Maki parameter ($\alpha_{1}$) by fitting the $\mu_{0}H_{\rm c2}(T)$ data to the following equation given by,
\begin{equation}
\mu_{0}H_{\rm c2}(T) = \frac{\mu_{0}H_{\rm c2}^{\rm orb}(0)}{\sqrt{1+\alpha_{1}^{2}}}
\{1 - (T/T_{\rm c})^{2} \}. \label{eq4}
\end{equation}
The fitting results are indicated by the dashed lines in Fig.~\ref{fig03}. 
The $\mu_{0}H_{\rm c2}^{\rm zero}(T)$ is in good agreement with the fitting line by Eq.~\ref{eq4}, 
but the $\mu_{0}H_{\rm c2}^{\rm onset}(T)$ and the $\mu_{0}H_{\rm c2}^{\rm mid}(T)$ are not satisfactory. 
The values of $\alpha_{1}$ are smaller than those of $\alpha_{0}$ for all defined critical fields. 
Since the Maki parameter is defined as $\alpha = \sqrt{2}H_{\rm c2}^{\rm orb}(0)/H_{\rm p}(0)$, 
it is expected that the decrease of $\alpha$ arises from the increase of $H_{\rm p}(0)$, suggesting the 
increase of $\Delta_{0}$ because the $\lambda_{\rm so}$ is nearly zero as reported by Fuchs $et~al$.~\cite{Fuchs}. Their report said that the effect of spin-orbit scattering on the upper critical field 
is expected to be rather weak for strongly disordered iron pnictide superconductors. 
In the ARPES (angle resolved photoemission spectroscopy) experiment on 
Ba$_{0.6}$K$_{0.4}$Fe$_{2}$As$_{2}$~\cite{ARPES}, 
a large superconductivity gap (2$\Delta/k_{\rm B}T_{\rm c} \sim 7.5$ at 7~K) was observed. 
The large superconductivity gap was also observed in strong-coupling high-$T_{\rm c}$ cuprates, 
such as Bi-2212~\cite{Bi2212} and YBCO~\cite{YBCO}.
These results suggest that 
the $\Delta_{0}$ of FeSe$_{0.25}$Te$_{0.75}$ is also larger than that for a conventional weak-coupling 
superconductor, indicating that the present compound is a strong-coupling superconductor. 
We have discussed the $\mu_{0}H_{\rm c2}$ by the extended WHH model, 
but this model is devised for one-band superconductors. 
Therefore, 
we hope that these studies on the upper critical fields will stimulate construction of the theories 
for multi-band superconductors.

In conclusion, 
we have investigated the upper critical fields $\mu_{0}H_{\rm c2}(T)$ of 
the 11-system iron-chalcogenide superconductor FeSe$_{0.25}$Te$_{0.75}$ ($T_{\rm c} \sim 14$~K). 
This compound shows very large initial slopes of $\mu_{0}H_{\rm c2}(T)$ at
$T_{\rm c}^{\rm onset}(0)$, $T_{\rm c}^{\rm mid}(0)$, and 
$T_{\rm c}^{\rm zero}(0)$ which are $-13.7$, $-10.1$, and $-6.9$~T/K, respectively.
The experimental upper critical fields of the present sample are smaller than those expected from the WHH model. 
The suppression of $\mu_{0}H_{\rm c2}$ at low temperature requires a large Maki parameter, 
but the parameter is smaller than that calculated for a weak-coupling superconductor, 
indicating a large superconducting gap as observed in strong-coupling high-$T_{\rm c}$ cuprates. 
Consequently, 
these results suggest that the iron-chalcogenide FeSe$_{0.25}$Te$_{0.75}$ is a strong-coupling superconductor. 

\section*{Acknowledgements} 
We thank Prof. Y. Ando, and Dr. K. Segawa (SANKEN, Osaka Univ.), and Dr. H. Eisaki (AIST) for fruitful discussions. 
This work was partly supported by 
a Grant-in-Aid for "Transformative Research-project on Iron Pnictides (TRIP)" 
from the Japan Science and Technology Agency (JST), 
a Grant-in-Aid for Scientific Research on priority Areas "High Field Spin Science in 100T" (No.451) 
from the Ministry of Education, Culture, Sports, Science and Technology (MEXT), Japan, and 
the Global COE Program 
(Core Research and Engineering of Advanced Materials-Interdisciplinary Education Center for 
Materials Science) (No. G10) from the MEXT, Japan.

\end{document}